\documentclass{aa}
\usepackage[varg]{txfonts} 
\usepackage{graphicx}  
\usepackage{dcolumn}   
\usepackage{bm}        
\usepackage{booktabs}
\usepackage{natbib}
\usepackage{rotating}
\usepackage{lscape}
\usepackage{xcolor}
\usepackage{siunitx}
\usepackage{makecell}
\usepackage{amsmath}
\usepackage[colorlinks=true,linkcolor=blue,citecolor=blue]{hyperref}

\bibpunct{(}{)}{;}{a}{}{,} 

\begin{document}

\title{Search for gravitational waves associated with high-energy
  neutrinos detected by IceCube during the third observing run of LIGO-Virgo}
\titlerunning{Search for GWs associated with IceCube neutrinos during O3}







\author{
  M. Vereecken\inst{\ref{inst1}}\corrauth{matthias.vereecken@ugent.be}
  \and M. Pracchia\inst{\ref{inst2}}\email{mpracchia@uliege.be}
  \and K. Merfeld\inst{\ref{inst3}}\email{karamerfeld5@gmail.com}
  \and I. Tosta e Melo\inst{\ref{inst4}}\email{iara.tosta.melo@dfa.unict.it}
  \and W. Javed\inst{\ref{inst5}}\email{JavedW@cardiff.ac.uk}
  \and P. J. Sutton\inst{\ref{inst5}}\email{SuttonPJ1@cardiff.ac.uk}
}
\authorrunning{Vereecken et al.}

\institute{
  Universiteit Gent, B-9000 Gent, Belgium\label{inst1}
  \and STAR Institute, Université de Liège, B-4000 Liège, Belgium\label{inst2}
  \and Johns Hopkins University, Baltimore, MD 21218, USA\label{inst3}
  \and University of Catania, Department of Physics and Astronomy, 95123 Catania CT, Italy\label{inst4}
  \and Cardiff University, Cardiff CF24 3AA, United Kingdom\label{inst5}
}

\date{Received ...; accepted ...}

  \abstract
   {}
{We search for generic gravitational-wave transients associated with high-energy neutrinos detected by
  the IceCube Neutrino Observatory during the third Observing Run (O3) of
  Advanced LIGO, Advanced Virgo, and KAGRA.} 
   {We perform an unmodeled, targeted search, which is sensitive
  to gravitational-wave signals weaker than those reported in real
  time or in the gravitional-wave transient catalog, and can thus
  uncover coincidences missed by typical neutrino follow-up searches.}
   {We find no statistically significant gravitational-wave signal and
  set lower bounds on the distance of possible gravitational-wave
  sources for different emission models.}
   {}


\maketitle
\nolinenumbers

\section{Introduction}

The first direct observation of gravitational waves, emitted by
the merger of two stellar-mass black
holes~\citep{LIGOScientific:2016aoc} in 2015, kicked off a new era in
multimessenger astronomy. With the recent release of version 5 of the
gravitational-wave transient catalog, GWTC-5~\citep{LIGOScientific:2026sit},
over 300 compact binary merger candidates have been
observed confidently.
When compact binary mergers involve at least one neutron star, they
are expected to emit radiation other than gravitational waves as
well. For example, short gamma-ray bursts (GRB) are typically expected
to be produced by the merger of a binary neutron star (BNS) system and
gravitational waves are emitted during the inspiral phase of the
system. In the most widely accepted scenario, tidally disrupted matter
from the system can form a torus around the newborn black hole or
neutron star formed during the merger, which can then power a
relativistic jet that can be observed if it is pointing towards
Earth. This picture was confirmed for the first time by the
unambiguous joint observation of short GRB (GRB~170817A) with a GW
signal from a BNS merger
(GW170817)~\citep{LIGOScientific:2017vwq,LIGOScientific:2017ync}. The
afterglow of the GRB was detected across the electromagnetic
spectrum~\citep{LIGOScientific:2017ync}. A large follow-up campaign
allowed to confidently observe for the first time the kilonova which
follows the GRB, and identify neutron star mergers as production sites
of heavy elements like gold and platinum. However, so far, only a
single multimessenger coincidence has been detected.

In addition to binary neutron star mergers, there is also the possibility of GRB emission in neutron star-black hole (NSBH) mergers. While there is currently no observational evidence that such systems can produce a GRB, under certain conditions the BH might tidally disrupt the NS while merging, creating an accretion disk that can power the GRB~\citep{Pannarale:2014rea}.

Long GRBs, on the other hand, are typically expected to be generated in collapse of a
massive star, although there is evidence some long GRBs are due to binary inspirals~\citep{Fynbo:2006mw,Gehrels:2006tk,Tanga:2017obk,Rastinejad:2022zbg,JWST:2023jqa,Levan:2023qdh}. In these so-called collapsars, the core of the massive star forms a stellar-mass black hole, and is surrounded by an accretion disk that powers the relativistic jet.

GRBs have long been proposed as sources of ultra-high
energy cosmic rays and, related to that, high-energy
neutrinos. However, no neutrinos were detected in coincidence with
GRB~170817A~\citep{ANTARES:2017bia}, or from other
GRBs (see e.g.~\citet{IceCube:2023woj}).
Despite that, GRBs remain a promising source of
high-energy neutrinos. In particular, the searches for neutrinos from GRBs have so
far focused on GRBs detected in GeV gamma rays. However, there are
models which favor neutrino production from GRBs which
are not bright in GeV gamma rays: neutrinos from core-collapse
supernovae with choked jets~\citep{Guetta:2019wpb}, or from choked jets
in low-luminosity GRBs~\citep{Senno:2015tsn}. Looking for gravitational
waves in coincidence with high-energy neutrinos forms an attractive
alternative way to detect such sources.

Observing coincident gravitational waves and neutrinos from a source
would present another major breakthrough in multimessenger astronomy,
since neutrinos can provide unique evidence of hadronic acceleration in
these sources. As such, there is a lot of interest in the possible
coincident detection of gravitational waves and
neutrinos~\citep{Ando:2012hna,Kimura:2018vvz,Murase:2019tjj}. This has
motivated many gravitational-wave follow-up searches in neutrinos from
MeV to hundreds of TeV by
Super-Kamiokande~\citep{Super-Kamiokande:2021dav},
ANTARES~\citep{ANTARES:2023wcj}, KM3NeT~\citep{KM3NeT:2023cdr}, and
IceCube~\citep{IceCube:2022mma,IceCube:2023atb}. These follow-up
searches use independently detected gravitational-wave candidates, and
perform a deep follow-up in neutrinos.

In this work, we present a search doing the inverse: a search for
gravitational waves from the direction of high-energy neutrino
candidates detected by IceCube reported in the IceCat-1
catalog~\citep{IceCube:2023agq}. In this way, if the neutrino is of
astrophysical origin, we know a source must exist in its direction. We
can then directly test whether this source emits gravitational
waves. Such searches are already performed within the LIGO-Virgo-KAGRA
collaboration for several other astrophysical source classes, e.g.\
GRBs~\citep{LIGOScientific:2016akj,LIGOScientific:2019obb,LIGOScientific:2020lst,LIGOScientific:2021iyk},
fast radio
bursts~\citep{LIGOScientific:2022jpr,LIGOScientific:2024avz}, and
magnetars~\citep{LIGOScientific:2019ccu,LIGOScientific:2022sts}. By
performing a coherent search in a small time window around the
neutrino trigger, and only in the direction of this trigger, targeted
searches can be more sensitive than all-sky searches. On average, the
gain in sensitivity, and thus distance, is around 20\%~\citep{Williamson:2014wma}. We perform a
generic search for transients, which is sensitive to a variety of
gravitational wave sources such as inspirals, accretion disk
instabilities, or generic unmodeled GW sources (also referred to as
``GW bursts''). A similar search for
gravitational waves in association with neutrino was carried out in
the previous generation of gravitational-wave detectors, together with
ANTARES~\citep{LIGOScientific:2012bvo}. In the decade since then, both
gravitational-wave and neutrino detectors have made significant
improvements in sensitivity.
The work here covers the third Observing Run of Advanced LIGO and
Advanced Virgo, which consisted of two parts: O3a (1 April 2019 15:00
UTC to 1 October 2019 15:00 UTC) and O3b (1 November 2019 15:00 UTC to
27 March 2020 17:00 UTC)~\citep{KAGRA:2023pio}.

In Sect.~\ref{sec:method}, we describe the method we used to search
for gravitational-wave transients. In Sect.~\ref{sec:models}, we present the
waveform models considered. In Sect.~\ref{sec:icecube}, we discuss
the neutrino events used for our search. In Sect.~\ref{sec:results},
we report our results. Finally, in Sect.~\ref{sec:conclusion} we
present our conclusion.
\section{Search method}
\label{sec:method}

We perform an unmodelled targeted search for generic GW transients
using the \texttt{X-Pipeline} software
package~\citep{Sutton:2009gi,Was:2012zq}. This type of analysis
searches for coherent excess power in multiple detectors, without
making any assumption on the GW signal morphology, as opposed to
modelled searches which uses matched filtering techniques with well
defined template banks of GW waveforms. Being a targeted search, data
is analyzed only in a small time window around an external
astrophysical trigger, usually referred to as the \textit{on-source}
window, and for arrival directions compatible with the sky
localization of the event. By limiting the search to a restricted time
window and sky position, trial factors from scanning over all times
and the entire sky are mitigated. Moreover, by taking into account the
antenna factor associated with the sky position external trigger, and
combining data from different detectors in a coherent way, the
statistical odds of a power excess due to random noise are reduced.

The search is performed in a time window of [-500,~+500]~s around an
external astrophysical neutrino trigger, motivated by an analysis of
typical GRB durations, without making strong
assumptions on which phase the neutrino emission is related
to~\citep{Baret:2011tk,von_Kienlin_2020}. We search for signals in a frequency range of
20--500~Hz, similar to the LVK GRB follow-up
search~\citep{LIGOScientific:2021iyk}. This range covers the region of
the most promising signals (like binary mergers) and where a generic
argument shows that we are sensitive out to the largest
distances~\citep{Sutton:2013ooa}.

\texttt{X-Pipeline} generates multiple coherent time-frequency maps of
the strain data, using short Fourier transforms with different time
resolutions $\Delta t = 2^n$, with $n$ being all the integer numbers
between $-7$ and $+1$.  Pixels of coherent excess energy forming
clusters in the spectrograms are considered as candidate GW events and
are ranked according to their total energy.  Coherent consistency
tests, based on correlated data between the detectors, allow the analysis to veto
events for which the relative energy in different detectors is not
consistent with the expectation from the event direction and is thus
most likely due to noise.  For events with a large sky position
uncertainty, the arrival time difference in different detectors
directions would cause a time shift between the GW signals in the
respective detectors data, resulting in the GW signal being likely
vetoed by a coherent consistency test~\citep{Was:2012zq}. For such
events, the analysis is performed over a discrete grid of sky
positions built within $2\sigma$ area of the statistical angular
uncertainty of the event.

The detection efficiency is optimized automatically by tuning the
event selection cuts using injections of simulated signal events for
different source classes (see Sect.~\ref{sec:models}). The surviving
event with the highest ranking statistic in the analysis time window
is then considered as the best gravitational-wave candidate.

The statistical significance of the gravitational-wave candidates is
evaluated by measuring their probability of being produce by
background fluctuations. The background noise is characterized within
at least $\sim 1.5$ hours of coincident data surrounding the trigger
time, outside of the on-source window. The so-called
\textit{off-source} data is split into chunks of the same length as
the on-source window, i.e. 1000~s, and each of those off-source trials
is analyzed in the same way as the on-source time window. By
time-sliding the data from different detectors relative to each other,
we increase the amount of background trials, assuming the different
detectors have uncorrelated instrumental noise. The loudest trigger
from each of the off-source trials is then used to build a
distribution of the loudest background triggers energy, from which we
can obtain the $p$-value for the on-source event.

Finally, the sensitivity of the search is determined by injecting
simulated astrophysical GW signals. We find the number of injections
which are recovered at higher significance than the loudest on-source event. The upper limit on amplitude for a given source class is then given by the amplitude for which we recover 90\% of all injections for that source class. This can be recast into an exclusion distance for that class, defined as the maximum distance at which 90\% of all injections are louder than the loudest trigger in the on-source window.


\section{Source models}
\label{sec:models}

We consider GRBs as the most likely sources to emit both
high-energy neutrinos and transient gravitational waves detectable by
ground-based interferometers. Therefore, we optimize our search for
different possible progenitors of GRBs, and obtain distance exclusion
limits for different classes of GW waveforms.



The first class of waveform models to which we tune our
search is the inspiral of compact objects. We consider neutron stars with masses
varying according to a normal distribution
$\mathcal{N}(1.4,~0.2)~M_\odot$ between $[1,3]~M_\odot$, and black
holes varying as $\mathcal{N}(10,~6)~M_\odot$ between
$[2,25]~M_\odot$.  In addition, for BNS systems the total mass is
constrained between $[2,~6]~M_\odot$, while for NSBH it is between
$[3,~25]~M_\odot$. Assuming collimated GRB jet emission, whose axis
coincides with the rotational axis of the binary system, the cosine of
the inclination angle is sampled between 0.866 and 1. However, for compact binary mergers, the gravitational waveform $h(t)$ can be accurately predicted. In that case, matched-filtering is the optimal method to search for such signals.

For what concerns GW waveforms associated with long GRBs, given
the high variability of waveforms predicted in hydrodynamical
simulations of these phenomena and the considerable uncertainty on the
physics involved, no strongly predictive waveforms are
available. Instead, we tune our search to two families of
representative generic gravitational-wave signals.

The first family of signals is circularly-polarized sine-Gaussian chirplets (CSG).  They cover the generic case where the gravitational-wave emission is due to
a rotational instability in the central engine, which creates a mass
quadrupole moment changing with time. The gravitational-wave signal is
modeled as a circularly polarized, gaussian-modulated sinusoid which
represents a generic burst signal.

Following~\citet{LIGOScientific:2016akj}, we have
\begin{equation}
\label{eq:csghh}
\begin{split}
\begin{bmatrix}h_+(t) \\ h_\times(t)\end{bmatrix} = & \frac{1}{r}
\sqrt{\frac{G}{c^3}\frac{E_{GW}}{f_0 Q}\frac{5}{4\pi^{3/2}}} \begin{bmatrix}(1+\cos^2\iota)\cos(2\pi f_0 t) \\ 2 \cos\iota
  \sin(2\pi f_0 t)\end{bmatrix}\\
&\times  \exp \left(-\frac{(2\pi f_0
  t)^2}{2Q^2}\right).
\end{split}
\end{equation}
with the signal frequency $f_0$ equal to twice the rotation frequency,
$t$ the time relative to peak time, and $Q$ the quality factor representing the
number of cycles for which the mass quadrupole moment is large.
The root sum square of the plus ($h_+$) and cross ($h_\times$)
polarizations, $h_{rss}$, is then given by
\begin{equation}
\label{eq:csghrss}
h_{rss} = \frac{1}{r}
\sqrt{\frac{G}{c^3}\frac{E_{GW}}{f_0^2}\frac{5}{2\pi^2}}
\end{equation}
We consider 4 cases of the CSG waveform: signals with frequencies
70~Hz, 100~Hz, 150~Hz, or 300~Hz, allowing $\cos\iota$ to vary between
0.996 and 1, and a time constant which is the inverse of the frequency (quality factor $Q$ of approximately 9)~\citep{LIGOScientific:2021iyk}.
In order to derive distance exclusion limits, we fix $E_\mathrm{GW}$
to $10^{-2}$~$M_\odot c^2$ as an optimistic estimate for the energy
released as gravitational waves from rotational instabilities
(see~\citet{LIGOScientific:2016akj} and references therein).

The second family of signals is the
Accretion Disk Instability
(ADI), 
following the model in~\citet{vanPutten:2001sw,vanPutten:2003hd}
which represents an extreme case of GW emission from collapsars.
In this model, the rapidly rotating Kerr black hole created during the core-collapse
forms an active nucleus, with a magnetically suspended torus accreting
onto the black hole. Gravitational waves are emitted from the
quadrupole moment created by clumps in the torus, and are powered by
the spin of the black hole through its magnetic connection with the
accretion disk. The model depends on the mass $M$ and dimensionless spin $\chi$ of the central black
hole, the fraction $\epsilon$ of disk mass that forms clumps, and the
mass of the accretion disk $M_\mathrm{disk}$. We use 5 different
combinations of these parameters, shown in Table~\ref{tab:adi}, that
cover different signal morphologies predicted by the model.

\begin{table*}[]
  \caption{Parameters used for the different accretion disk
	instability models~\citep{vanPutten:2001sw,vanPutten:2003hd} considered in our search,
    following~\citet{LIGOScientific:2016akj}. The mass and
    dimensionless spin of the central black hole are denoted by $M$
    and $\chi$ respectively, $\epsilon$ is the fraction of the disk mass
    that forms clumps, and the accretion disk mass is always equal to
    1.5~$M_\odot$. The resulting duration, frequency range, and total
    radiated energy are also shown in the table.}
  \label{tab:adi}
  \centering
  \begin{tabular}{lcccccccc}
    \toprule
    Label &
    $M$ ($M_\odot$) &
    $\chi$ & $\epsilon$ &
    
    Dur. (s) &
    Freq. (Hz) &
    $E_\mathrm{GW}$ ($M_\odot c^2$)\\
    \midrule
ADI-A & 5  & 0.3  & 0.05   & 39  & 135--166 & 0.02 \\
ADI-B & 10 & 0.95 & 0.2    & 9.4 & 110--209 & 0.22 \\
ADI-C & 10 & 0.95 & 0.04   & 236 & 130--251 & 0.25 \\
ADI-D & 3  & 0.7  & 0.035  & 142 & 119--173 & 0.02 \\
ADI-E & 8  & 0.99 & 0.065  & 76  & 111--260 & 0.17 \\
    \bottomrule
\end{tabular}
\end{table*}
\section{IceCube alerts}
\label{sec:icecube}

We look for gravitational waves coming from the direction of promising
neutrino candidates detected by the IceCube neutrino
observatory~\citep{IceCube:2016zyt}. IceCube has instrumented a cubic
kilometer of ice at the South Pole. High-energy neutrinos interact in
the ice, producing an electron, muon, or hadronic debris. The
Cherenkov radiation produced by these as they propagate through the
ice can then be detected.  In addition to astrophysical neutrinos,
IceCube also detects a large background from atmospheric muons and, to
a lesser extent, atmospheric neutrinos, both created in cosmic ray air
shower. By performing different event selections, based on the
topology, energy, and direction of observed neutrino candidates, the
IceCube Collaboration produces several datasets varying in purity and
type of neutrinos. 

While there are several publicly available datasets, we choose to make use of
the first IceCube Event Catalog of Alert Tracks
(IceCat-1)~\citep{IceCube:2023agq}. This catalog is based on the real-time
alerts sent out by IceCube over GCN since June 2019,
which allow for electromagnetic follow-up of neutrino candidates with
a high probability of astrophysical origin. At the moment, there are
two IceCube-only alert streams available depending on the topology of
the neutrino event: track alerts due to charged current interactions
of muon neutrinos in the ice, and cascade alerts due to charged
current interactions of electron neutrinos or neutral current
interactions of any flavor neutrino in the ice. However, during Observing Run 3
of LIGO-Virgo-KAGRA, only track alerts were available. IceCat-1 contains
all the track alerts sent out over GCN, as well as events from an
archival reanalysis of data before June 2019 that would have been sent
out over GCN. It also uses a more accurate reconstruction of the
neutrino events, and a veto for events which are likely due to cosmic
rays.

Track alerts are classified in two categories: Bronze and Gold alerts. These
alerts are sent out with a threshold such that the average probability to be of
astrophysical origin of the bronze (gold) sample is 30\% (50\%). This
probability is based on the candidate neutrino energy and declination: higher
energy neutrinos are more likely to be of astrophysical origin due to the
steeply falling atmospheric neutrino spectrum. Because of the significant
background from atmospheric muons in the southern atmosphere, the energy
threshold depends on declination.


The median 90\% uncertainty contour area of candidate neutrinos in IceCat-1 is 6.2 deg$^2$~\citep{IceCube:2025uzh}.
These uncertainties are provided as HEALPix
maps~\citep{Gorski:2004by}.
However, since our gravitational-wave analysis uses Gaussian error
regions by default, we circularize the HEALPix error region to
estimate the angular error by averaging the distance from the most
probable direction to the 1 sigma contour. We checked the obtained estimate with
the angular errors provided in the GCN alerts to ensure
consistency. These errors are shown in Table~\ref{tab:fulltable}.

We select all the alerts in IceCat-1 that have no cosmic-ray veto. In addition, we include
\texttt{133781\_21701751} (Bronze),
which does not appear in IceCat-1, but is still analyzed based on the GCN alert~\citep{GCNBronze}.
This gives 10 Bronze alerts and 7 Gold alerts in O3a, as well as 6
Bronze alerts and 2 Gold alerts in O3b. However, the O3a events
\texttt{132792\_60166398} (Bronze) and
\texttt{132910\_57145925} (Gold)
have no interferometer data available and therefore cannot be
analyzed.
In total, our final list of analyzed candidates consists of 15 Bronze and 8
Gold alerts.


\section{Results}
\label{sec:results}

\begin{figure}
\includegraphics[width=\columnwidth]{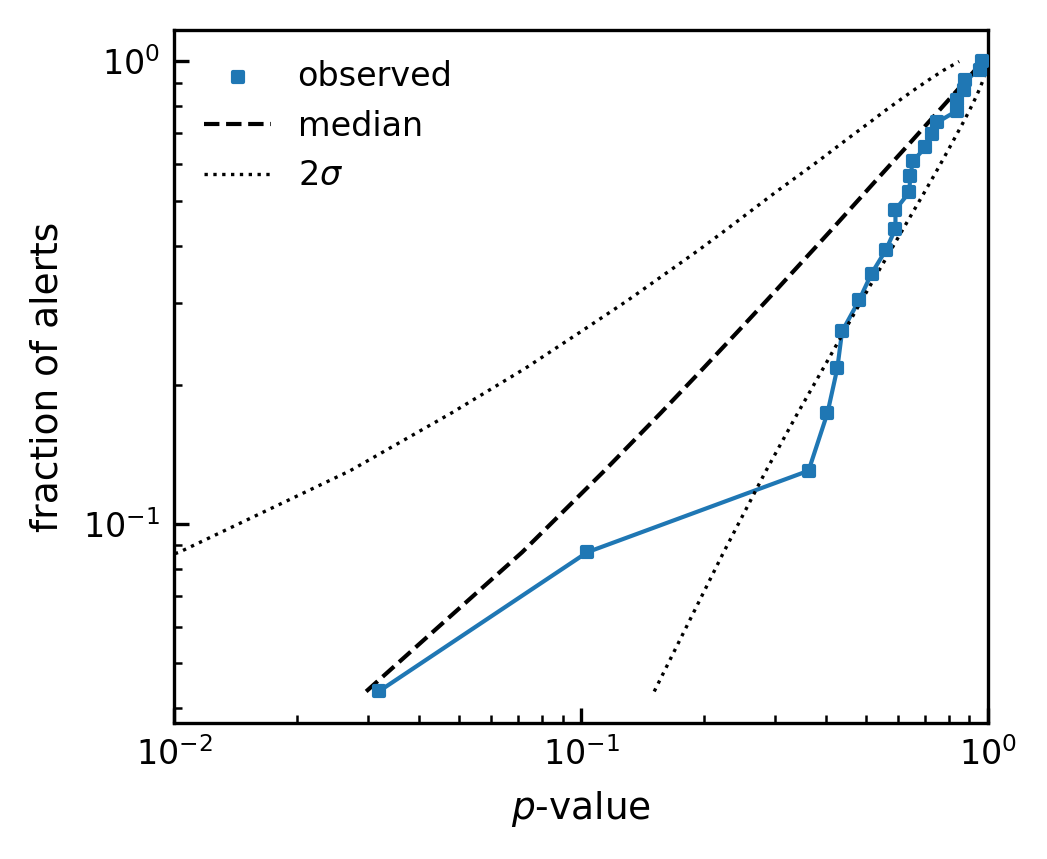}
\caption{Cumulative $p$-value distribution for the analyzed events
  (blue). The dashed black line shows the median expectation obtained
  from $p$-values drawn from a uniform distribution, while the gray
  lines indicate the $2\sigma$-band around this median.}
\label{fig:cumpvalue}
\end{figure}

We show the results from the unmodeled targeted search for
gravitational waves for the 15 Bronze and 8 Gold neutrino events detected by
IceCube during Observing Run 3 that have interferometer data
available in Table~\ref{tab:fulltable}.
The cumulative distribution of $p$-values is shown in
Fig.~\ref{fig:cumpvalue}, along with the expectation from the null
hypothesis, i.e. no gravitational waves associated with high-energy
neutrinos, by drawing $p$-values from a uniform distribution. The
lowest p-value we observe, associated with event
\texttt{132577\_42662743}, is $p=0.032$. Seeing one such $p$-value
in 23 analyses is well within the expected range. We therefore
find no evidence of gravitational waves associated with any of the
high-energy neutrino events analyzed. If there was a subset of
neutrino alerts with weak gravitational wave signals, this
distribution would show an excess of low $p$-vales. Since such excess
is not observed, we also find no evidence of gravitational waves
associated with high-energy neutrino on a population level.

\begin{figure*}
\includegraphics[width=\textwidth]{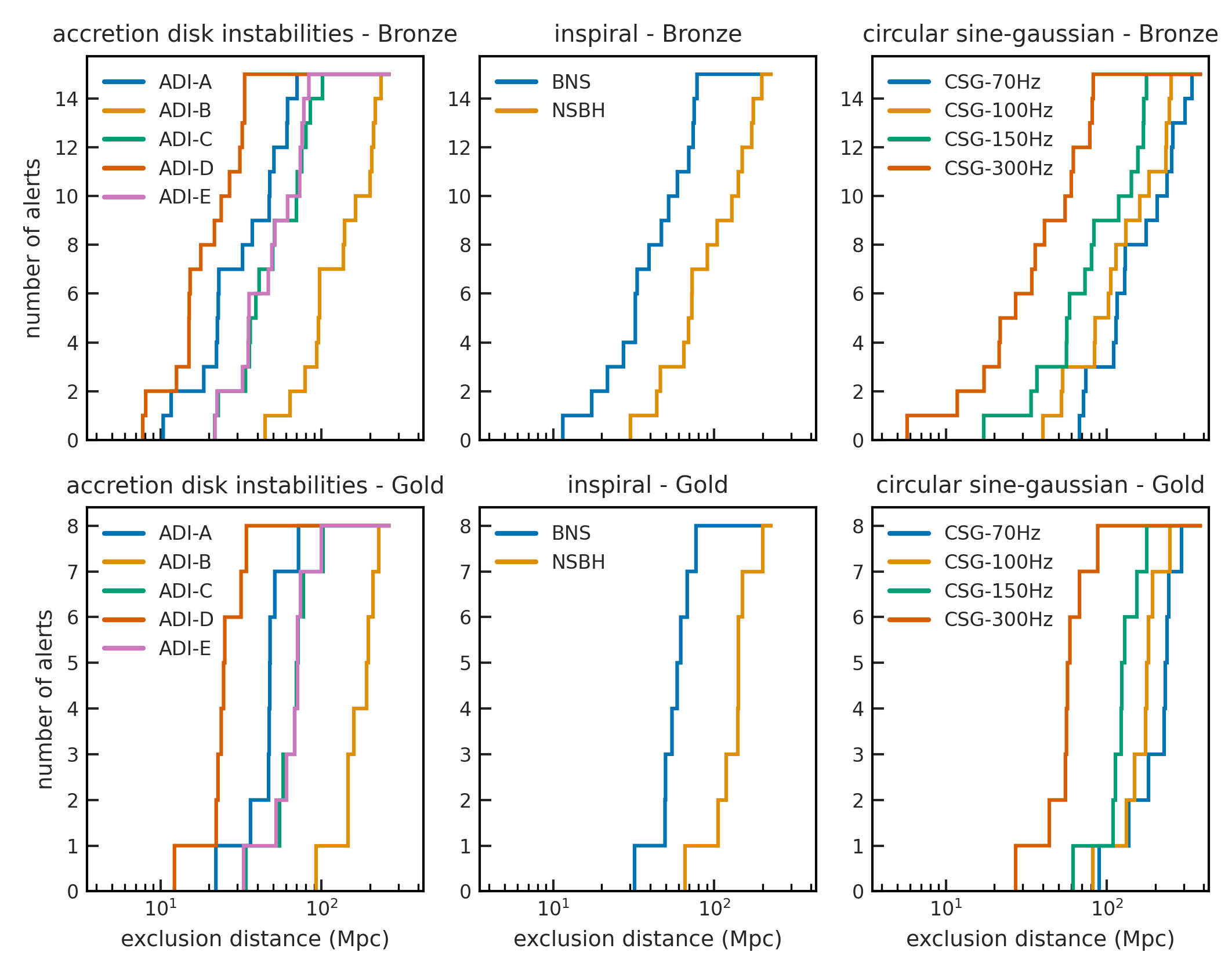}
\caption{Exclusion distances at 90\% confidence level obtained for the different waveform
  classes for Bronze (top row) and Gold (bottom row) events.}
\label{fig:excldist}
\end{figure*}

Given this null observation, we can derive distance exclusion limits
on possible gravitational-wave sources associated with the high-energy
neutrinos, with the understanding it only applies to an event if it is of astrophysical origin. We show 90\% exclusion distance, i.e.\ the distance for
which we expect to detect the gravitational wave signal 90\% of the
time, in Fig.~\ref{fig:excldist} for the different waveforms that
were tested, separated by event type. In addition, the median
exclusion distance for each waveform and event type are shown in
Table~\ref{tab:meddist}. We can see that the distance exclusion limits
are slightly better for Gold alerts. However, the sensitivity depends
on a combination of antenna pattern of the interferometers at the time
and direction of the neutrino event, the number of interferometers
online at the time, the level of detector noise, and the neutrino angular
error. For the results shown, the antenna factor and the number of
detectors online are the dominant factors. Therefore, the difference
in sensitivity between Bronze and Gold events is purely
accidental. Still, we show them separately, because Gold events are
more likely to be of astrophysical origin.

For binary neutron star mergers, the 90\% exclusion distance is around
30--60~Mpc. As expected, this is less than the single-detector
sensitivities to binary neutron star mergers quoted by LVK for the O3
LIGO detectors of 100--140 Mpc~\citep{KAGRA:2021vkt}, since
matched-filtering is more sensitive than an unmodeled
search in the case of compact binary coalescences. Likewise, the 90\% exclusion distance for neutron star-black
hole mergers is around 70-140~Mpc, lower than that expected from
matched-filtering searches.
On the other hand, an unmodeled search can uniquely detect sources
other than inspirals. For both accretion disk instabilities and circular
sine-gaussians, the sensitivity depends on the chosen model, varying
between tens of Mpc to a few hundred Mpc.
For accretion disk instabilities, the models with higher
$E_\mathrm{GW}$ and shorter duration have the strongest exclusion limits (see
Table~\ref{tab:adi}). For CSG models, lower frequency models have the
strongest exclusion limits, since they lead to higher $h_{rss}$ for
equal $E_\mathrm{GW}$ (see Eq.~\ref{eq:csghrss}).

\begin{table}[]
\caption{Median 90\% exclusion distances for the Bronze and Gold
  events for each of the different source classes considered.}
\label{tab:meddist}
\centering
\begin{tabular}{lcc}
\toprule
 & Bronze (Mpc) & Gold (Mpc) \\
Waveform &  &  \\
\midrule
ADI-A & 32 & 48 \\
ADI-B & 137 & 175 \\
ADI-C & 50 & 69 \\
ADI-D & 18 & 24 \\
ADI-E & 49 & 70 \\
BNS & 39 & 57 \\
NSBH & 91 & 141 \\
CSG-70Hz & 130 & 229 \\
CSG-100Hz & 114 & 177 \\
CSG-150Hz & 81 & 124 \\
CSG-300Hz & 36 & 57 \\
\bottomrule
\end{tabular}
\end{table}

\section{Conclusions}
\label{sec:conclusion}
We performed a targeted, unmodeled search for gravitational wave transients
coincident with high-energy neutrino events from IceCat-1 detected by
IceCube, using data collected by LIGO and Virgo during Observing run
O3. Typical sources for such coincident emission are short GRBs, expected mostly from the merger of a binary neutron star system, or long GRBs, expected mostly from the core
collapse of a massive star. However, since our search is unmodeled, it
is sensitive to any generic source of gravitational waves.

We found no statistical evidence of gravitational waves coincident
with high-energy neutrinos. As such, we present exclusion limits for a
variety of waveform models, ranging from about 10 Mpc up to 300
Mpc. We considered both Bronze and Gold neutrino events which have, on average over the alert sample, 
30\% and 50\% probability of astrophysical origin respectively.

The current analysis does not include cascade-like events created by neutral
current interaction of muon neutrinos, or any interaction from electron and tau
neutrinos. Since such events do not suffer from the atmospherical muon
background, any identified cascade event is very likely to be of astrophysical
origin~\citep{gcn_cascade_alerts}. However, cascade alerts were not yet
available from IceCube during Observing run O3 of LVK. For observing run O4,
such alerts are available and are a promising additional target. Moreover, an
updated IceCat-2 is expected to be released soon, with significantly improved
angular uncertainties~\citep{IceCube:2025uzh}. Finally, in the future, a
targeted modeled search can improve constraints for the case of compact binary
mergers.

\begin{acknowledgements}
The authors thank Biswajit Banerjee for constructive comments that helped
improve this manuscript.
  The authors are grateful for computational resources provided by the
  LIGO Laboratory and supported by National Science Foundation Grants
  PHY-0757058 and PHY-0823459.  
MV was supported as postdoctoral research fundamental research by the
Fonds Wetenschappelijk Onderzoek - Vlaanderen under grant No. 1298826N.
MP was supported by the Fonds de la Recherche Scientifique - FNRS, Belgium, under grant No. 4.4501
This research has made use of data or software obtained from the
Gravitational Wave Open Science Center (gwosc.org), a service of the
LIGO Scientific Collaboration, the Virgo Collaboration, and
KAGRA. This material is based upon work supported by NSF's LIGO
Laboratory which is a major facility fully funded by the National
Science Foundation, as well as the Science and Technology Facilities
Council (STFC) of the United Kingdom, the Max-Planck-Society (MPS),
and the State of Niedersachsen/Germany for support of the construction
of Advanced LIGO and construction and operation of the GEO600
detector. Additional support for Advanced LIGO was provided by the
Australian Research Council. Virgo is funded, through the European
Gravitational Observatory (EGO), by the French Centre National de
Recherche Scientifique (CNRS), the Italian Istituto Nazionale di
Fisica Nucleare (INFN) and the Dutch Nikhef, with contributions by
institutions from Belgium, Germany, Greece, Hungary, Ireland, Japan,
Monaco, Poland, Portugal, Spain. KAGRA is supported by Ministry of
Education, Culture, Sports, Science and Technology (MEXT), Japan
Society for the Promotion of Science (JSPS) in Japan; National
Research Foundation (NRF) and Ministry of Science and ICT (MSIT) in
Korea; Academia Sinica (AS) and National Science and Technology
Council (NSTC) in Taiwan.
\end{acknowledgements}

\bibliographystyle{aa}
\bibliography{references}

@article{LIGOScientific:2012bvo,
    author = "Adrian-Martinez, S. and Al Samarai, I. and Albert, A. and others",
    collaboration = "LIGO Scientific, VIRGO",
    title = "{A First Search for coincident Gravitational Waves and High Energy Neutrinos using LIGO, Virgo and ANTARES data from 2007}",
    eprint = "1205.3018",
    archivePrefix = "arXiv",
    primaryClass = "astro-ph.HE",
    reportNumber = "LIGO-P1200006",
    doi = "10.1088/1475-7516/2013/06/008",
    journal = "JCAP",
    volume = "06",
    pages = "008",
    year = "2013"
}

@article{LIGOScientific:2026sit,
	author = "Abac, A. G. and Abe, A. and Abouelfettouh, I. and others",
	collaboration = "LIGO Scientific, VIRGO, KAGRA",
	title = "{GWTC-5.0: An Introduction to Version 5.0 of the Gravitational-Wave Transient Catalog}",
	eprint = "2605.27223",
	archivePrefix = "arXiv",
	primaryClass = "gr-qc",
	reportNumber = "LIGO-P2500701",
	month = "5",
	year = "2026"
}

@article{IceCube:2023atb,
    author = "Abbasi, R. and Ackermann, M. and Adams, J. and others",
    collaboration = "IceCube",
    title = "{A Search for IceCube Sub-TeV Neutrinos Correlated with Gravitational-wave Events Detected By LIGO/Virgo}",
    eprint = "2303.15970",
    archivePrefix = "arXiv",
    primaryClass = "astro-ph.HE",
    doi = "10.3847/1538-4357/aceefc",
    journal = "Astrophys. J.",
    volume = "959",
    number = "2",
    pages = "96",
    year = "2023",
    note = "[Erratum: Astrophys.J. 971, 192 (2024)]"
}

@article{Senno:2015tsn,
    author = "Senno, N. and Murase, K. and Meszaros, P.",
    title = "{Choked Jets and Low-Luminosity Gamma-Ray Bursts as Hidden Neutrino Sources}",
    eprint = "1512.08513",
    archivePrefix = "arXiv",
    primaryClass = "astro-ph.HE",
    doi = "10.1103/PhysRevD.93.083003",
    journal = "Phys. Rev. D",
    volume = "93",
    number = "8",
    pages = "083003",
    year = "2016"
}

@article{Guetta:2019wpb,
    author = "Guetta, D. and Rahin, R. and Bartos, I. and Della Valle, M.",
    title = "{Constraining the fraction of core-collapse supernovae harbouring choked jets with high-energy neutrinos}",
    eprint = "1906.07399",
    archivePrefix = "arXiv",
    primaryClass = "astro-ph.HE",
    doi = "10.1093/mnras/stz3245",
    journal = "Mon. Not. Roy. Astron. Soc.",
    volume = "492",
    number = "1",
    pages = "843--847",
    year = "2020"
}

@article{LIGOScientific:2017vwq,
    author = "Abbott, B. P. and Abbott, R. and Abbott, T. D and others",
    collaboration = "LIGO Scientific, Virgo",
    title = "{GW170817: Observation of Gravitational Waves from a Binary Neutron Star Inspiral}",
    eprint = "1710.05832",
    archivePrefix = "arXiv",
    primaryClass = "gr-qc",
    reportNumber = "LIGO-P170817",
    doi = "10.1103/PhysRevLett.119.161101",
    journal = "Phys. Rev. Lett.",
    volume = "119",
    number = "16",
    pages = "161101",
    year = "2017"
}

@article{Murase:2019tjj,
    author = "Murase, K. and Bartos, I.",
    title = "{High-Energy Multimessenger Transient Astrophysics}",
    eprint = "1907.12506",
    archivePrefix = "arXiv",
    primaryClass = "astro-ph.HE",
    doi = "10.1146/annurev-nucl-101918-023510",
    journal = "Ann. Rev. Nucl. Part. Sci.",
    volume = "69",
    pages = "477--506",
    year = "2019"
}

@article{IceCube:2023agq,
    author = "Abbasi, R. and Ackermann, M. and Adams, J.  and others",
    collaboration = "IceCube",
    title = "{IceCat-1: The IceCube Event Catalog of Alert Tracks}",
    eprint = "2304.01174",
    archivePrefix = "arXiv",
    primaryClass = "astro-ph.HE",
    doi = "10.3847/1538-4365/acfa95",
    journal = "Astrophys. J. Suppl.",
    volume = "269",
    number = "1",
    pages = "25",
    year = "2023"
}

@article{IceCube:2022mma,
    author = "Abbasi, R. and Ackermann, M. and Adams, J. and others",
    collaboration = "IceCube",
    title = "{IceCube Search for Neutrinos Coincident with Gravitational Wave Events from LIGO/Virgo Run O3}",
    eprint = "2208.09532",
    archivePrefix = "arXiv",
    primaryClass = "astro-ph.HE",
    doi = "10.3847/1538-4357/aca5fc",
    journal = "Astrophys. J.",
    volume = "944",
    number = "1",
    pages = "80",
    year = "2023"
}

@article{LIGOScientific:2017ync,
    author = "Abbott, B. P. and Abbott, R. and Abbott, T. D. and others",
    collaboration = "LIGO Scientific, Virgo, Fermi GBM, INTEGRAL, IceCube, AstroSat Cadmium Zinc Telluride Imager Team, IPN, Insight-Hxmt, ANTARES, Swift, AGILE Team, 1M2H Team, Dark Energy Camera GW-EM, DES, DLT40, GRAWITA, Fermi-LAT, ATCA, ASKAP, Las Cumbres Observatory Group, OzGrav, DWF (Deeper Wider Faster Program), AST3, CAASTRO, VINROUGE, MASTER, J-GEM, GROWTH, JAGWAR, CaltechNRAO, TTU-NRAO, NuSTAR, Pan-STARRS, MAXI Team, TZAC Consortium, KU, Nordic Optical Telescope, ePESSTO, GROND, Texas Tech University, SALT Group, TOROS, BOOTES, MWA, CALET, IKI-GW Follow-up, H.E.S.S., LOFAR, LWA, HAWC, Pierre Auger, ALMA, Euro VLBI Team, Pi of Sky, Chandra Team at McGill University, DFN, ATLAS Telescopes, High Time Resolution Universe Survey, RIMAS, RATIR, SKA South Africa/MeerKAT",
    title = "{Multi-messenger Observations of a Binary Neutron Star Merger}",
    eprint = "1710.05833",
    archivePrefix = "arXiv",
    primaryClass = "astro-ph.HE",
    reportNumber = "LIGO-P1700294, VIR-0802A-17, FERMILAB-PUB-17-478-A-AE-CD",
    doi = "10.3847/2041-8213/aa91c9",
    journal = "Astrophys. J. Lett.",
    volume = "848",
    number = "2",
    pages = "L12",
    year = "2017"
}

@article{LIGOScientific:2016aoc,
    author = "Abbott, B. P. and Abbott, R. and Abbott, T. D. and others",
    collaboration = "LIGO Scientific, Virgo",
    title = "{Observation of Gravitational Waves from a Binary Black Hole Merger}",
    eprint = "1602.03837",
    archivePrefix = "arXiv",
    primaryClass = "gr-qc",
    reportNumber = "LIGO-P150914",
    doi = "10.1103/PhysRevLett.116.061102",
    journal = "Phys. Rev. Lett.",
    volume = "116",
    number = "6",
    pages = "061102",
    year = "2016"
}

@article{Was:2012zq,
    author = "Was, M. and Sutton, P. J. and Jones, G. and Leonor, I.",
    title = "{Performance of an externally triggered gravitational-wave burst search}",
    eprint = "1201.5599",
    archivePrefix = "arXiv",
    primaryClass = "gr-qc",
    reportNumber = "LIGO-P1100135",
    doi = "10.1103/PhysRevD.86.022003",
    journal = "Phys. Rev. D",
    volume = "86",
    pages = "022003",
    year = "2012"
}

@article{IceCube:2023woj,
    author = "Abbasi, R. and Ackermann, M. and Adams, J. and others",
    collaboration = "IceCube",
    title = "{Search for 10\textendash{}1000 GeV Neutrinos from Gamma-Ray Bursts with IceCube}",
    eprint = "2312.11515",
    archivePrefix = "arXiv",
    primaryClass = "astro-ph.HE",
    doi = "10.3847/1538-4357/ad220b",
    journal = "Astrophys. J.",
    volume = "964",
    number = "2",
    pages = "126",
    year = "2024"
}

@article{LIGOScientific:2022jpr,
    author = "Abbott, R. and Abbott, T. D. and Acernese, F. and others",
    collaboration = "LIGO Scientific, VIRGO, KAGRA, CHIME/FRB",
    title = "{Search for Gravitational Waves Associated with Fast Radio Bursts Detected by CHIME/FRB during the LIGO\textendash{}Virgo Observing Run O3a}",
    eprint = "2203.12038",
    archivePrefix = "arXiv",
    primaryClass = "astro-ph.HE",
    reportNumber = "P2100124",
    doi = "10.3847/1538-4357/acd770",
    journal = "Astrophys. J.",
    volume = "955",
    number = "2",
    pages = "155",
    year = "2023"
}

@article{LIGOScientific:2021iyk,
    author = "Abbott, R. and Abbott, T. D. and Acernese, F. and others",
    collaboration = "LIGO Scientific, VIRGO, KAGRA",
    title = "{Search for Gravitational Waves Associated with Gamma-Ray Bursts Detected by Fermi and Swift during the LIGO\textendash{}Virgo Run O3b}",
    eprint = "2111.03608",
    archivePrefix = "arXiv",
    primaryClass = "astro-ph.HE",
    reportNumber = "P2100091",
    doi = "10.3847/1538-4357/ac532b",
    journal = "Astrophys. J.",
    volume = "928",
    number = "2",
    pages = "186",
    year = "2022"
}

@article{LIGOScientific:2022sts,
    author = "Abbott, R. and Abe, H. and Acernese, F. and others",
    collaboration = "LIGO Scientific, Virgo,, KAGRA, VIRGO",
    title = "{Search for Gravitational-wave Transients Associated with Magnetar Bursts in Advanced LIGO and Advanced Virgo Data from the Third Observing Run}",
    eprint = "2210.10931",
    archivePrefix = "arXiv",
    primaryClass = "astro-ph.HE",
    reportNumber = "LIGO-P2100387",
    doi = "10.3847/1538-4357/ad27d3",
    journal = "Astrophys. J.",
    volume = "966",
    number = "1",
    pages = "137",
    year = "2024"
}

@article{ANTARES:2017bia,
    author = "Albert, A. and André, M. and Anghinolfi, M. and others",
    collaboration = "ANTARES, IceCube, Pierre Auger, LIGO Scientific, Virgo",
    title = "{Search for High-energy Neutrinos from Binary Neutron Star Merger GW170817 with ANTARES, IceCube, and the Pierre Auger Observatory}",
    eprint = "1710.05839",
    archivePrefix = "arXiv",
    primaryClass = "astro-ph.HE",
    reportNumber = "LIGO-P1700344, FERMILAB-PUB-17-471-ND-TD, LIGO-P1700344",
    doi = "10.3847/2041-8213/aa9aed",
    journal = "Astrophys. J. Lett.",
    volume = "850",
    number = "2",
    pages = "L35",
    year = "2017"
}

@article{ANTARES:2023wcj,
    author = "Albert, A. and Alves, S. and André, M. and others",
    collaboration = "ANTARES",
    title = "{Search for neutrino counterparts to the gravitational wave sources from LIGO/Virgo O3 run with the ANTARES detector}",
    eprint = "2302.07723",
    archivePrefix = "arXiv",
    primaryClass = "astro-ph.HE",
    doi = "10.1088/1475-7516/2023/04/004",
    journal = "JCAP",
    volume = "04",
    pages = "004",
    year = "2023"
}

@article{Super-Kamiokande:2021dav,
    author = "Abe, K. and Bronner, C. and Hayato, Y. and others",
    collaboration = "Super-Kamiokande",
    title = "{Search for neutrinos in coincidence with gravitational wave events from the LIGO-Virgo O3a Observing Run with the Super-Kamiokande detector}",
    eprint = "2104.09196",
    archivePrefix = "arXiv",
    primaryClass = "astro-ph.HE",
    doi = "10.3847/1538-4357/ac0d5a",
    journal = "Astrophys. J.",
    volume = "918",
    number = "2",
    pages = "78",
    year = "2021"
}

@article{KM3NeT:2023cdr,
    author = "Aiello, S. and Albert, A. and Alves Garre, S. and others",
    collaboration = "KM3NeT",
    title = "{Searches for neutrino counterparts of gravitational waves from the LIGO/Virgo third observing run with KM3NeT}",
    eprint = "2311.03804",
    archivePrefix = "arXiv",
    primaryClass = "astro-ph.HE",
    doi = "10.1088/1475-7516/2024/04/026",
    journal = "JCAP",
    volume = "04",
    pages = "026",
    year = "2024"
}

@article{Kimura:2018vvz,
    author = "Kimura, Shigeo S. and Murase, K. and Bartos, I. and others",
    title = "{Transejecta high-energy neutrino emission from binary neutron star mergers}",
    eprint = "1805.11613",
    archivePrefix = "arXiv",
    primaryClass = "astro-ph.HE",
    doi = "10.1103/PhysRevD.98.043020",
    journal = "Phys. Rev. D",
    volume = "98",
    number = "4",
    pages = "043020",
    year = "2018"
}

@article{Sutton:2009gi,
    author = "Sutton, P. J. and Jones, G. and Chatterji, S. and others",
    title = "{X-Pipeline: An Analysis package for autonomous gravitational-wave burst searches}",
    eprint = "0908.3665",
    archivePrefix = "arXiv",
    primaryClass = "gr-qc",
    doi = "10.1088/1367-2630/12/5/053034",
    journal = "New J. Phys.",
    volume = "12",
    pages = "053034",
    year = "2010"
}

@article{Ando:2012hna,
    author = "Ando, S. and Baret, B. and Bartos, I. and others",
    title = "{Colloquium: Multimessenger astronomy with gravitational waves and high-energy neutrinos}",
    eprint = "1203.5192",
    archivePrefix = "arXiv",
    primaryClass = "astro-ph.HE",
    doi = "10.1103/RevModPhys.85.1401",
    journal = "Rev. Mod. Phys.",
    volume = "85",
    number = "4",
    pages = "1401--1420",
    year = "2013"
}

@article{Baret:2011tk,
    author = "Baret, B. and Bartos, I. and Bouhou, B. and others",
    title = "{Bounding the Time Delay between High-energy Neutrinos and Gravitational-wave Transients from Gamma-ray Bursts}",
    eprint = "1101.4669",
    archivePrefix = "arXiv",
    primaryClass = "astro-ph.HE",
    doi = "10.1016/j.astropartphys.2011.04.001",
    journal = "Astropart. Phys.",
    volume = "35",
    pages = "1--7",
    year = "2011"
}

@article{LIGOScientific:2016akj,
    author = "Abbott, B. P. and Abbott, R. and Abbott, T. D. and others",
    collaboration = "LIGO Scientific, Virgo, IPN",
    title = "{Search for Gravitational Waves Associated with Gamma-Ray Bursts During the First Advanced LIGO Observing Run and Implications for the Origin of GRB 150906B}",
    eprint = "1611.07947",
    archivePrefix = "arXiv",
    primaryClass = "astro-ph.HE",
    reportNumber = "P1600298",
    doi = "10.3847/1538-4357/aa6c47",
    journal = "Astrophys. J.",
    volume = "841",
    number = "2",
    pages = "89",
    year = "2017"
}

@article{LIGOScientific:2019obb,
    author = "Abbott, B. P. and Abbott, R. and Abbott, T. D. and others",
    collaboration = "LIGO Scientific, Virgo, IPN",
    title = "{Search for gravitational-wave signals associated with gamma-ray bursts during the second observing run of Advanced LIGO and Advanced Virgo}",
    eprint = "1907.01443",
    archivePrefix = "arXiv",
    primaryClass = "astro-ph.HE",
    reportNumber = "LIGO-P1900034",
    doi = "10.3847/1538-4357/ab4b48",
    journal = "Astrophys. J.",
    volume = "886",
    pages = "75",
    year = "2019"
}

@article{LIGOScientific:2020lst,
    author = "Abbott, R. and Abbott, T. D. and Abraham, S. and others",
    collaboration = "LIGO Scientific, Virgo",
    title = "{Search for Gravitational Waves Associated with Gamma-Ray Bursts Detected by Fermi and Swift During the LIGO-Virgo Run O3a}",
    eprint = "2010.14550",
    archivePrefix = "arXiv",
    primaryClass = "astro-ph.HE",
    reportNumber = "LIGO-P2000040",
    doi = "10.3847/1538-4357/abee15",
    journal = "Astrophys. J.",
    volume = "915",
    number = "2",
    pages = "86",
    year = "2021"
}

@article{KAGRA:2021vkt,
    author = "Abbott, R. and Abbott, T. D. and Acernese, F. and others",
    collaboration = "KAGRA, VIRGO, LIGO Scientific",
    title = "{GWTC-3: Compact Binary Coalescences Observed by LIGO and Virgo during the Second Part of the Third Observing Run}",
    eprint = "2111.03606",
    archivePrefix = "arXiv",
    primaryClass = "gr-qc",
    reportNumber = "LIGO-P2000318",
    doi = "10.1103/PhysRevX.13.041039",
    journal = "Phys. Rev. X",
    volume = "13",
    number = "4",
    pages = "041039",
    year = "2023"
}

@article{Gorski:2004by,
    author = "G{\'o}rski, K. M. and Hivon, E. and Banday, A. J. and others",
    title = "{HEALPix - A Framework for high resolution discretization, and fast analysis of data distributed on the sphere}",
    eprint = "astro-ph/0409513",
    archivePrefix = "arXiv",
    doi = "10.1086/427976",
    journal = "Astrophys. J.",
    volume = "622",
    pages = "759--771",
    year = "2005"
}

@article{Rastinejad:2022zbg,
    author = "Rastinejad, J. C. and Gompertz, B. P. and Levan, A. J. and others",
    title = "{A kilonova following a long-duration gamma-ray burst at 350 Mpc}",
    eprint = "2204.10864",
    archivePrefix = "arXiv",
    primaryClass = "astro-ph.HE",
    doi = "10.1038/s41586-022-05390-w",
    journal = "Nature",
    volume = "612",
    number = "7939",
    pages = "223--227",
    year = "2022"
}

@article{Tanga:2017obk,
    author = {Tanga, M. and Kr{\"u}hler, T. and Schady, P. and others},
    title = "{The environment of the SN-less GRB 111005A at z = 0.0133}",
    eprint = "1708.06270",
    archivePrefix = "arXiv",
    primaryClass = "astro-ph.GA",
    doi = "10.1051/0004-6361/201731799",
    journal = "Astron. Astrophys.",
    volume = "615",
    pages = "A136",
    year = "2018"
}

@article{Fynbo:2006mw,
    author = "Fynbo, J. P. U. and Watson, D. and Th{\"o}ne, C. C. and others",
    title = "{No supernovae from two nearby long gamma ray bursts}",
    eprint = "astro-ph/0608313",
    archivePrefix = "arXiv",
    doi = "10.1038/nature05375",
    journal = "Nature",
    volume = "444",
    pages = "1047--1049",
    year = "2006"
}

@ARTICLE{Gehrels:2006tk,
       author = {{Gehrels}, N. and {Norris}, J.~P. and {Barthelmy}, S.~D. and others},
        title = "{A new {\ensuremath{\gamma}}-ray burst classification scheme from GRB060614}",
      journal = {\nat},
         year = 2006,
        month = dec,
       volume = {444},
       number = {7122},
        pages = {1044-1046},
          doi = {10.1038/nature05376},
archivePrefix = {arXiv},
       eprint = {astro-ph/0610635},
 primaryClass = {astro-ph},
       adsurl = {https://ui.adsabs.harvard.edu/abs/2006Natur.444.1044G}
}

@article{JWST:2023jqa,
    author = "Levan, A. J. and Gompertz, B. P. and Salafia, O. S. and others",
    collaboration = "JWST",
    title = "{Heavy-element production in a compact object merger observed by JWST}",
    eprint = "2307.02098",
    archivePrefix = "arXiv",
    primaryClass = "astro-ph.HE",
    doi = "10.1038/s41586-023-06759-1",
    journal = "Nature",
    volume = "626",
    number = "8000",
    pages = "737--741",
    year = "2024"
}

@article{Levan:2023qdh,
    author = "Levan, A. and Malesani, D. and Gompertz, B. P. and others",
    title = "{A long-duration gamma-ray burst of dynamical origin from the nucleus of an ancient galaxy}",
    eprint = "2303.12912",
    archivePrefix = "arXiv",
    primaryClass = "astro-ph.HE",
    doi = "10.1038/s41550-023-01998-8",
    journal = "Nature Astron.",
    volume = "7",
    number = "8",
    pages = "976--985",
    year = "2023"
}

@article{vanPutten:2001sw,
    author = "van Putten, M. H. P. M.",
    title = "{Proposed source of gravitational radiation from a torus around a black hole}",
    eprint = "astro-ph/0107007",
    archivePrefix = "arXiv",
    doi = "10.1103/PhysRevLett.87.091101",
    journal = "Phys. Rev. Lett.",
    volume = "87",
    pages = "091101",
    year = "2001"
}

@article{vanPutten:2003hd,
    author = "van Putten, M. H. P M. and Levinson, A. and Lee, H. K. and others",
    title = "{Gravitational radiation from gamma-ray bursts as observational opportunities for LIGO and VIRGO}",
    eprint = "gr-qc/0308016",
    archivePrefix = "arXiv",
    reportNumber = "LIGO-P030041-00-D",
    doi = "10.1103/PhysRevD.69.044007",
    journal = "Phys. Rev. D",
    volume = "69",
    pages = "044007",
    year = "2004"
}

@article{Sutton:2013ooa,
    author = "Sutton, P. J.",
    title = "{A Rule of Thumb for the Detectability of Gravitational-Wave Bursts}",
    eprint = "1304.0210",
    archivePrefix = "arXiv",
    primaryClass = "gr-qc",
    reportNumber = "LIGO-P1000041-V3",
    month = "3",
    year = "2013"
}

@article{KAGRA:2023pio,
    author = "Abbott, R. and Abe, H. and Acernese, F. and others",
    collaboration = "KAGRA, VIRGO, LIGO Scientific",
    title = "{Open Data from the Third Observing Run of LIGO, Virgo, KAGRA and GEO}",
    eprint = "2302.03676",
    archivePrefix = "arXiv",
    primaryClass = "gr-qc",
    reportNumber = "LIGO-P2200316",
    doi = "10.3847/1538-4365/acdc9f",
    journal = "Astrophys. J. Suppl.",
    volume = "267",
    number = "2",
    pages = "29",
    year = "2023"
}

@ARTICLE{GCNBronze,
       author = {{IceCube Collaboration}},
        title = "{IceCube-200227A: IceCube observation of a high-energy neutrino candidate event}",
      journal = {GRB Coordinates Network},
         year = 2020,
        month = feb,
       volume = {27235},
        pages = {1},
       adsurl = {https://ui.adsabs.harvard.edu/abs/2020GCN.27235....1I}
}

@article{Pannarale:2014rea,
    author = "Pannarale, F. and Ohme, F.",
    title = "{Prospects for joint gravitational-wave and electromagnetic observations of neutron-star--black-hole coalescing binaries}",
    eprint = "1406.6057",
    archivePrefix = "arXiv",
    primaryClass = "gr-qc",
    reportNumber = "LIGO-P1400096",
    doi = "10.1088/2041-8205/791/1/L7",
    journal = "Astrophys. J. Lett.",
    volume = "791",
    pages = "L7",
    year = "2014"
}

@article{IceCube:2016zyt,
    author = "Aartsen, M. G. and Ackermann, M. and Adams, J. and others",
    collaboration = "IceCube",
    title = "{The IceCube Neutrino Observatory: Instrumentation and Online Systems}",
    eprint = "1612.05093",
    archivePrefix = "arXiv",
    primaryClass = "astro-ph.IM",
    doi = "10.1088/1748-0221/12/03/P03012",
    journal = "JINST",
    volume = "12",
    number = "03",
    pages = "P03012",
    year = "2017",
    note = "[Erratum: JINST 19, E05001 (2024)]"
}

@article{LIGOScientific:2019ccu,
    author = "Abbott, B. P. and Abbott, R. and Abbott, T. D. and others",
    collaboration = "LIGO Scientific, Virgo",
    title = "{Search for Transient Gravitational-wave Signals Associated with Magnetar Bursts during Advanced LIGO{\textquoteright}s Second Observing Run}",
    eprint = "1902.01557",
    archivePrefix = "arXiv",
    primaryClass = "astro-ph.HE",
    reportNumber = "LIGO-P1800165",
    doi = "10.3847/1538-4357/ab0e15",
    journal = "Astrophys. J.",
    volume = "874",
    number = "2",
    pages = "163",
    year = "2019"
}

@article{LIGOScientific:2024avz,
    author = "Abac, A. G. and Abbott, R. and Abouelfettouh, I. and others",
    collaboration = "LIGO Scientific, KAGRA, VIRGO",
    title = "{A Search Using GEO600 for Gravitational Waves Coincident with Fast Radio Bursts from SGR 1935+2154}",
    eprint = "2410.09151",
    archivePrefix = "arXiv",
    primaryClass = "astro-ph.HE",
    reportNumber = "LIGO-P2400192",
    doi = "10.3847/1538-4357/ad8de0",
    journal = "Astrophys. J.",
    volume = "977",
    number = "2",
    pages = "255",
    year = "2024"
}

@article{Williamson:2014wma,
    author = "Williamson, A. R. and Biwer, C. and Fairhurst, S. and others",
    title = "{Improved methods for detecting gravitational waves associated with short gamma-ray bursts}",
    eprint = "1410.6042",
    archivePrefix = "arXiv",
    primaryClass = "gr-qc",
    reportNumber = "LIGO-P1400044",
    doi = "10.1103/PhysRevD.90.122004",
    journal = "Phys. Rev. D",
    volume = "90",
    number = "12",
    pages = "122004",
    year = "2014"
}

@article{von_Kienlin_2020,
	title={The Fourth Fermi-GBM Gamma-Ray Burst Catalog: A Decade of Data},
	volume={893},
	ISSN={1538-4357},
	url={http://dx.doi.org/10.3847/1538-4357/ab7a18},
	DOI={10.3847/1538-4357/ab7a18},
	number={1},
	journal={The Astrophysical Journal},
	publisher={American Astronomical Society},
	author={von Kienlin, A. and Meegan, C. A. and Paciesas, W. S. and others},
	year={2020},
	month=Apr, pages={46} }

@techreport{gcn_cascade_alerts,
  author      = "{{IceCube Collaboration}}",
  title       =  "{High Energy Neutrino Cascade Alerts}",
  institution = {NASA},
  year = "2020",
  note        = {Accessed: 2025-11-24},
  url         = {http://gcn.gsfc.nasa.gov/doc/High_Energy_Neutrino_Cascade_Alerts.pdf}
}

@article{IceCube:2025uzh,
    author = "Abbasi, Rasha and others",
    collaboration = "IceCube",
    title = "{IceCat-2: Updated IceCube Event Catalog of Alert Tracks}",
    eprint = "2507.06176",
    archivePrefix = "arXiv",
    primaryClass = "astro-ph.HE",
    reportNumber = "PoS(ICRC2025)1224",
    doi = "10.22323/1.501.1224",
    journal = "PoS",
    volume = "ICRC2025",
    pages = "1224",
    year = "2025"
}


\longtab[1]{
  \begin{landscape}
    \small
\setlength{\tabcolsep}{3pt}
\begin{longtable}{llcccllcccccccccccc}
\caption{Neutrino events analyzed in this work, and their $90\%$ exclusion
distance for each source model tested. Exclusion distances are
expressed in Mpc.} \label{tab:fulltable} \\
\hline \hline
 & Type & {R.A. ($^\circ$)} & {Dec. ($^\circ$)} & {\(\sigma\) (\(^\circ\))} & Run & Network & $p$-value & {ADI-A} & {ADI-B} & {ADI-C} & {ADI-D} & {ADI-E} & {BNS} & {NSBH} & {\makecell[c]{CSG\\70Hz}} & {\makecell[c]{CSG\\100Hz}} & {\makecell[c]{CSG\\150Hz}} & {\makecell[c]{CSG\\300Hz}} \\
Alert &  &  &  &  &  &  &  &  &  &  &  &  &  &  &  &  &  &  \\
\hline
132427\_70353420 & Bronze & 310.61 & 12.22 & 1.35 & O3a & H1L1V1 & 0.10 & 47 & 163 & 70 & 24 & 62 & 52 & 129 & 206 & 161 & 119 & 55 \\
132437\_16335312 & Bronze & 219.33 & 11.72 & 0.43 & O3a & H1L1V1 & 0.84 & 61 & 211 & 80 & 32 & 76 & 74 & 172 & 306 & 246 & 169 & 82 \\
132437\_67132865 & Bronze & 245.57 & 21.98 & 0.58 & O3a & H1V1 & 0.40 & 23 & 97 & 36 & 15 & 36 & 32 & 69 & 116 & 85 & 56 & 22 \\
132443\_12627143 & Bronze & 154.86 & 5.27 & 1.20 & O3a & H1V1 & 0.36 & 10 & 64 & 23 & 8 & 22 & 17 & 44 & 72 & 53 & 37 & 17 \\
132465\_3856549 & Bronze & 166.90 & 17.39 & 1.23 & O3a & H1L1 & 0.59 & 71 & 234 & 102 & 33 & 84 & 79 & 198 & 340 & 250 & 177 & 81 \\
132518\_766165 & Bronze & 65.17 & -37.26 & 0.61 & O3a & L1V1 & 0.65 & 23 & 79 & 41 & 15 & 47 & 22 & 46 & 75 & 53 & 34 & 12 \\
132768\_5390846 & Bronze & 29.12 & 84.56 & 2.26 & O3a & H1L1V1 & 0.52 & 51 & 205 & 76 & 31 & 74 & 70 & 149 & 254 & 234 & 156 & 60 \\
132792\_60166398 & Bronze & 161.81 & 26.90 & 1.17 & --- & --- & --- & --- & --- & --- & --- & --- & --- & --- & --- & --- & --- & --- \\
132814\_44222682 & Bronze & 76.64 & 12.75 & 2.25 & O3a & H1L1 & 0.44 & 37 & 139 & 51 & 22 & 51 & 39 & 91 & 129 & 114 & 83 & 41 \\
132974\_67924813 & Bronze & 148.54 & 1.45 & 0.83 & O3a & H1L1V1 & 0.97 & 62 & 217 & 86 & 33 & 78 & 75 & 175 & 258 & 236 & 170 & 79 \\
132508\_42419327 & Gold & 120.19 & 6.43 & 0.32 & O3a & H1L1V1 & 0.73 & 36 & 147 & 58 & 23 & 52 & 50 & 119 & 182 & 150 & 110 & 44 \\
132577\_42662743 & Gold & 127.88 & 12.60 & 0.29 & O3a & H1L1V1 & 0.03 & 51 & 209 & 78 & 32 & 74 & 62 & 142 & 231 & 175 & 130 & 59 \\
132684\_5635104 & Gold & 312.19 & 26.57 & 0.34 & O3a & H1L1V1 & 0.96 & 47 & 146 & 55 & 22 & 61 & 50 & 106 & 138 & 133 & 114 & 55 \\
132707\_54984442 & Gold & 343.52 & 10.28 & 1.24 & O3a & H1L1 & 0.64 & 72 & 228 & 102 & 34 & 101 & 77 & 201 & 292 & 248 & 178 & 88 \\
132910\_57145925 & Gold & 226.14 & 10.77 & 0.66 & --- & --- & --- & --- & --- & --- & --- & --- & --- & --- & --- & --- & --- & --- \\
133091\_81419 & Gold & 167.30 & -22.27 & 1.32 & O3a & H1L1V1 & 0.48 & 48 & 191 & 68 & 25 & 71 & 68 & 150 & 243 & 193 & 154 & 68 \\
133092\_52499868 & Gold & 5.71 & -1.53 & 0.50 & O3a & H1L1V1 & 0.56 & 47 & 159 & 70 & 24 & 68 & 55 & 140 & 237 & 178 & 123 & 57 \\
133348\_80807014 & Bronze & 27.03 & 0.07 & 0.91 & O3b & L1V1 & 0.84 & 22 & 96 & 36 & 15 & 35 & 32 & 73 & 130 & 103 & 56 & 21 \\
133394\_27261780 & Bronze & 80.16 & 2.87 & 0.84 & O3b & L1V1 & 0.70 & 32 & 137 & 50 & 18 & 49 & 47 & 104 & 176 & 132 & 81 & 34 \\
133433\_29047901 & Bronze & 286.83 & 58.45 & 0.76 & O3b & H1L1V1 & 0.64 & 19 & 98 & 39 & 15 & 35 & 33 & 73 & 114 & 106 & 73 & 36 \\
133572\_82361476 & Bronze & 48.47 & 20.11 & 2.13 & O3b & H1L1V1 & 0.43 & 23 & 93 & 34 & 13 & 32 & 27 & 65 & 111 & 84 & 59 & 27 \\
133634\_1410505 & Bronze & 116.02 & 29.18 & 0.40 & O3b & H1L1 & 0.59 & 48 & 201 & 71 & 27 & 73 & 59 & 142 & 237 & 183 & 143 & 62 \\
133781\_21701751 & Bronze & 348.26 & 21.32 & 0.24 & O3b & H1V1 & 0.75 & 12 & 45 & 22 & 8 & 23 & 11 & 30 & 68 & 40 & 17 & 6 \\
133331\_47828126 & Gold & 229.31 & 3.77 & 1.81 & O3b & H1L1V1 & 0.87 & 22 & 93 & 34 & 12 & 33 & 32 & 66 & 90 & 82 & 62 & 27 \\
133609\_37927131 & Gold & 165.45 & 11.80 & 1.32 & O3b & H1L1V1 & 0.88 & 48 & 196 & 72 & 25 & 71 & 59 & 142 & 228 & 181 & 124 & 56 \\
  \hline
\end{longtable}
\end{landscape}
}

\end{document}